\documentclass[amsmath,amssymb,aps,prl,reprint,floatfix,superscriptaddress]{revtex4-2}

\usepackage{graphicx}
\usepackage{dcolumn}
\usepackage{xcolor}
\usepackage{bm}
\usepackage[version=4]{mhchem}
\usepackage[breaklinks,colorlinks,linkcolor={blue}, %
            citecolor={blue},urlcolor={blue}]{hyperref}
\begin{document}

\title{Discovery of T center-like quantum defects in silicon}
\author{Yihuang Xiong} 
\affiliation{Thayer School of Engineering, Dartmouth College, Hanover, New Hampshire 03755, USA}
\author{Jiongzhi Zheng}
\affiliation{Thayer School of Engineering, Dartmouth College, Hanover, New Hampshire 03755, USA}
\author{Shay McBride}
\affiliation{Thayer School of Engineering, Dartmouth College, Hanover, New Hampshire 03755, USA}
\author{Xueyue Zhang}
\affiliation{Department of Electrical Engineering and Computer Sciences, University of California, Berkeley, California 94720, USA}
\affiliation{Department of Physics, University of California, Berkeley, California 94720, USA}
\author{Sin\'ead M.\ Griffin}
\affiliation{Materials Sciences Division, Lawrence Berkeley National Laboratory, Berkeley, California 94720, USA}
\affiliation{Molecular Foundry Division, Lawrence Berkeley National Laboratory, Berkeley, California 94720, USA}

\author{Geoffroy Hautier} 
\affiliation{Thayer School of Engineering, Dartmouth College, Hanover, New Hampshire 03755, USA}
\date{\today}

\begin{abstract}
Quantum technologies would benefit from the development of high performance quantum defects acting as single-photon emitters or spin-photon interface. Finding such a quantum defect in silicon is especially appealing in view of its favorable spin bath and high processability. While some color centers in silicon have been emerging in quantum applications, there is still a need to search and develop new high performance quantum emitters. Searching a high-throughput computational database of more than 22,000 charged complex defects in silicon, we identify a series of defects formed by a group III element combined with carbon ((A-C)$\rm _{Si}$ with A=B,Al,Ga,In,Tl) and substituting on a silicon site. These defects are analogous structurally, electronically and chemically to the well-known T center in silicon ((C-C-H)$\rm_{Si}$) and their optical properties are mainly driven by an unpaired electron in a carbon $p$ orbital. They all emit in the telecom and some of these color centers show improved properties compared to the T center in terms of computed radiative lifetime or emission efficiency. We also show that the synthesis of hydrogenated T center-like defects followed by a dehydrogenation annealing step could be an efficient way of synthesis.  All the T center-like defects show a higher symmetry than the T center making them easier to align with magnetic fields. Our work motivates further studies on the synthesis and control of this new family of quantum defects, and also demonstrates the use of high-throughput computational screening to detect new complex quantum defects.

\end{abstract}

\maketitle

\section{Introduction}

Quantum Information Science (QIS) hold the promise of revolutionizing the way we communicate, sense and compute \cite{Wolfowicz2021,Yan2021,Atature2018,Lukin2020}. Qubits have been explored in various physical systems, including superconducting circuits, trapped ions, quantum dots or point defects \cite{Arute.Nat.2019,Harty.PRL.2014,Warburton.Nat.Mat.2013,Robledo.Nat.2011}. Quantum defects can act as an ``artificial atom'' in the semiconductor host ``vacuum'' with their spin state being controlled by light. These defects, often called spin-photon interfaces, have been proposed as the fundamental building blocks of quantum communication and distributed modular quantum computing\cite{Ruf.JAP.2021,Simmons.PRXQuantum.2024}. While the nitrogen-vacancy (NV) center in diamond has been the poster child for quantum defects, it is far from perfect, suffering from issues such as high spectral diffusion and low efficiency (low Debye-Waller factor). Diamond is also not easy to process and nanopattern. Consequently, many efforts have been directed towards the search for new quantum defects in alternative hosts. Silicon has recently gained interest as it is the most mature semiconductor offering endless possibilities in nanofabrication and scalability\cite{Yan.APL.2021}. In addition, silicon has a  favorable nuclei spin bath that should lead to high spin coherence times (T$_2$)\cite{Kanai.PNAS.2022}. There is a recent surge of interest from the QIS community in re-investigating known color centers in silicon that emits in the telecom. A series of these color centers (e.g., the G center, the C center, and the T center) are now being synthesized, integrated in devices and used for quantum operations\cite{Komza2022arXiv,Redjem2020,Udvarhelyi.npj.2022,Dhaliah2022,Bergeron2020, Neumann2008, Higginbottom2022, Johnston2023arXiv,Islam.NanoLetter.2024, DeAbreu.OE.2023}.


These recent successes call for a more systematic search of new quantum defects in silicon. Here, we used a high-throughput computational screening approach to discover a new class of complex defects. These new defects are based on a group III element bound to carbon and substituting on a silicon site (e.g., (B-C)$\rm _{Si}$, (Al-C)$\rm _{Si}$ and (Ga-C)$\rm _{Si}$). These color centers are good candidates for spin-photon interfaces as they are paramagnetic and offer emission in the telecom as well as other favorable optical properties such as brightness and efficiency. We relate their atomistic and electronic structure to the T center (C-C-H)$\rm _{Si}$ which is currently the prominent paramagnetic quantum defect in silicon\cite{Dhaliah2022,Bergeron2020, Higginbottom2022, Johnston2023arXiv,Islam.NanoLetter.2024, DeAbreu.OE.2023}. Because of this analogy, we call these new color centers T center-like defects and rationalize their similarities to the T center based on chemical arguments. Notably, some of these T center-like defects can outperform the T center in terms of optical properties, sensitivity to magnetic field alignment, or yield during synthesis.

\section{Results}

\subsection{High-throughput screening and electronic structures}

\begin{figure*}[t]
 	\centering
 	\includegraphics[width=0.9\textwidth]{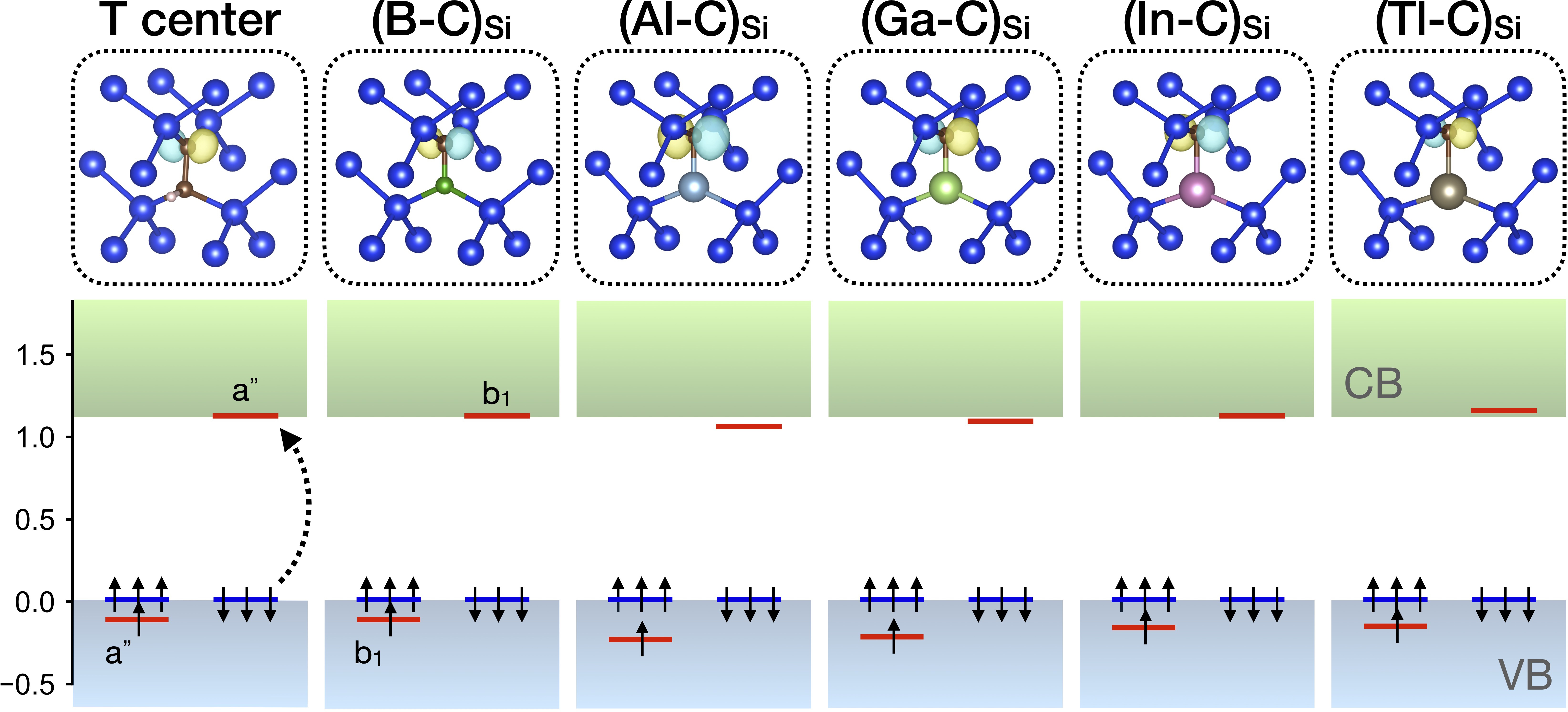}    
 	\caption{Defect atomic structures and single-particle Kohn-Sham levels computed at full HSE levels. T center and the T-like centers (acceptor-carbon pair) show similar electronic structures with a$^{''}$ and b$_1$ orbitals close to the CBM. The wavefunctions of the defect level show their $p$ orbital character from C dangling bonds.}
 	\label{Fig.2 HSE electronic}
\end{figure*}

We used a high-throughput computational screening approach to identify promising quantum defects in silicon. We start off by constructing a electronic-structure database of defects. Our previous work explored interstitial and substitutional defects in silicon \cite{Xiong2023}. In this study, we focus on substitutional-interstitial complex defects. Each of these complex defect is initialized by combining a substitutional defect with a nearest-neighbor interstitial either hexagonal or tetrahedral. We considered a set of 56 elements that are readily ion-implantable. By combining the stable charge states of simple defects \cite{Xiong2023}, we constructed a database comprising over 22,000 complex defects in various charge states.

Using this database, we searched for promising spin-photon interfaces that show charge stability for a range of the Fermi level within the band gap, non-singlet ground state, bright emission (transition dipole moment $>$2 D), and emission within the telecom wavelength. At the screening level, we approximated the emission wavelength using the Kohn-Sham energy difference at the single-shot HSE level (HSE$_0$). We refer to the method section and our previous work for detailed screening procedures \cite{Xiong2023}. We identify a class of five complex defects, all in neutral charge state and exhibiting a doublet ground state. We follow up by performing full self-consistent HSE computations on these defects and show the defect Kohn-Sham levels and their structures in Figure~\ref{Fig.2 HSE electronic}, with the T center ((C-C-H)$_{\rm Si}$) shown as a reference. The T center is excited through a defect-bound exciton where a hole in the silicon valence band is formed while the unoccupied a'' defect state is filled \cite{Dhaliah2022}. The excited state can be described by a delocalized hole attracted to a negatively charged (C-C-H)$_{\rm Si}$. All our identified defects are composed of a group III (B, Al, Ga, In, Tl)-carbon pair that occupies one Si site and exhibit a similar atomic structures to the T center (see Figure~\ref{Fig.2 HSE electronic}). Their electronic structure is also very similar with only one unoccupied defect level in the band gap that lies close to the conduction band. We refer to these as T center-like defects and they are also excited through a defect-bound exciton. Defect-bound exciton quantum defects have disadvantages (e.g., low operating temperature, weak brightness) compared to the two-level defect structures such as the NV center in diamond. However, in small band gap materials like silicon, they are often more easy to find when telecom emission is targeted \cite{Xiong.mqt.2024, Xiong2023}. Bound-exciton defects are amenable to quantum operations as demonstrated by the T center.

Similarly to the T center, the defect level in the band gap for the T-like defects originates from the $p$ orbital of the dangling bond on the carbon atom, as shown by the wavefunction in the defect structures (see Figure~\ref{Fig.2 HSE electronic}) \cite{Dhaliah2022}. This similarity can be understood from a simple chemical picture: in the T center, each C$^{4+}$ contains four dangling bonds and only one of the carbon is saturated by the hydrogen, thus the unsaturated dangling bond on the other carbon atom give rise to the defect level. This bond order can also be satisfied by replacing the C-H with a ground III acceptor (A$^{3+}$), which forms the T-like defects in our case. The T center has a symmetry of $C_{1h}$ due to the presence of the hydrogen atom, while the T-like defects have a higher symmetry of $C_{2v}$.

\begin{figure}[t]
 	\centering
 	\includegraphics[width=0.45\textwidth]{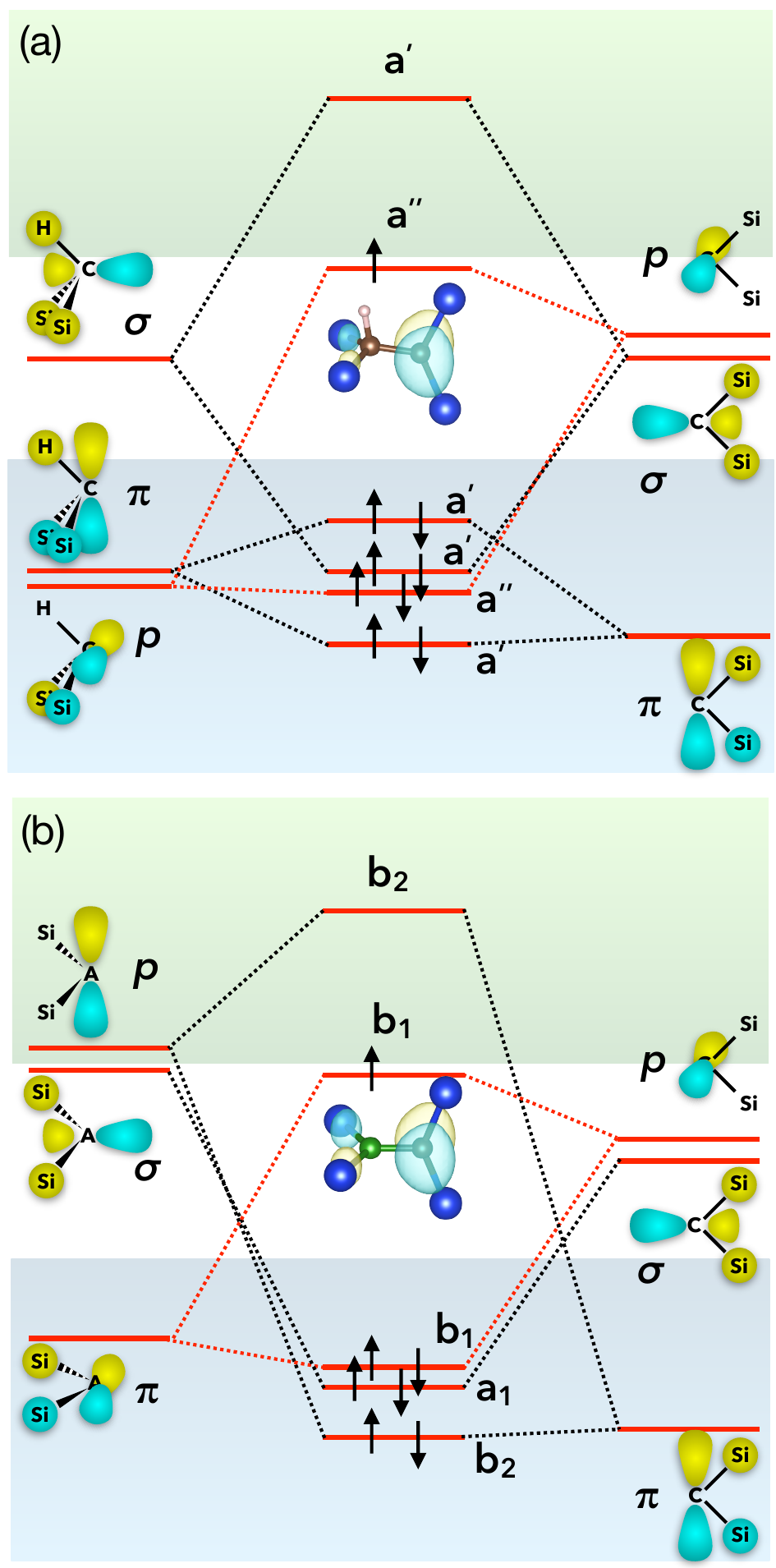}    
 	\caption{Molecular orbital diagrams of (a) T center the (b) T-like center (B-C)$_{\rm Si}$ in this case). In either case the frontier orbital is dominate by the unsaturated -C fragment, results in the nearly constant a$^{''}$/b$_1$ orbital in those defects.
 	}
 	\label{Fig.3 MO}
\end{figure}

The electronic structures of the T and T-like centers can be further understood using a fragment molecular orbital (MO) picture as shown in Figure~\ref{Fig.3 MO}. For the T center, the interaction between the $p$ orbital of -CH and the $p$ orbital of the -C gives rise to the anti-bonding a$^{''}$ orbital close to the conduction band minimum (CBM) (see Figure ~\ref{Fig.3 MO}a). We note this orbital is only occupied by one electron and the spin splitting is not captured by the MO schematics. Similarly, the interaction between the $\pi$ orbital of the acceptor and $p$ orbital of the -C fragment results in the frontier orbital lying close to the CBM, this is now denoted as $b_1$ due to the $C_{2v}$ symmetry of the T-like defects (see Figure ~\ref{Fig.3 MO}b). In both case the $p$ orbital from the unsaturated -C fragment dominates the frontier orbital, thus different acceptor atoms barely influence the a$^{''}$/b$_1$ level positions. We further corroborated this by substituting both saturated and unsaturated C with Ge and Si as discussed in Supplementary Note 1. While replacing C in -CH barely alters the defect levels, substituting unsaturated C reduced the splitting between a$^{''}$/b$_1$ for T/T-like centers cross the spin channels. This results in undesired defect level in the spin-up channel (see Supplementary Note 1). This highlights the critical role on optical properties played by unpaired electron in the $p$ orbital in the T center and T center-like defects. This a''/b$_1$ p orbital is perfectly positioned close to the conduction band allowing emission at an appropriate wavelength in the telecom through a defect-bound exciton process.

\begin{figure}[h]
 	\centering
 	\includegraphics[width=0.5\textwidth]{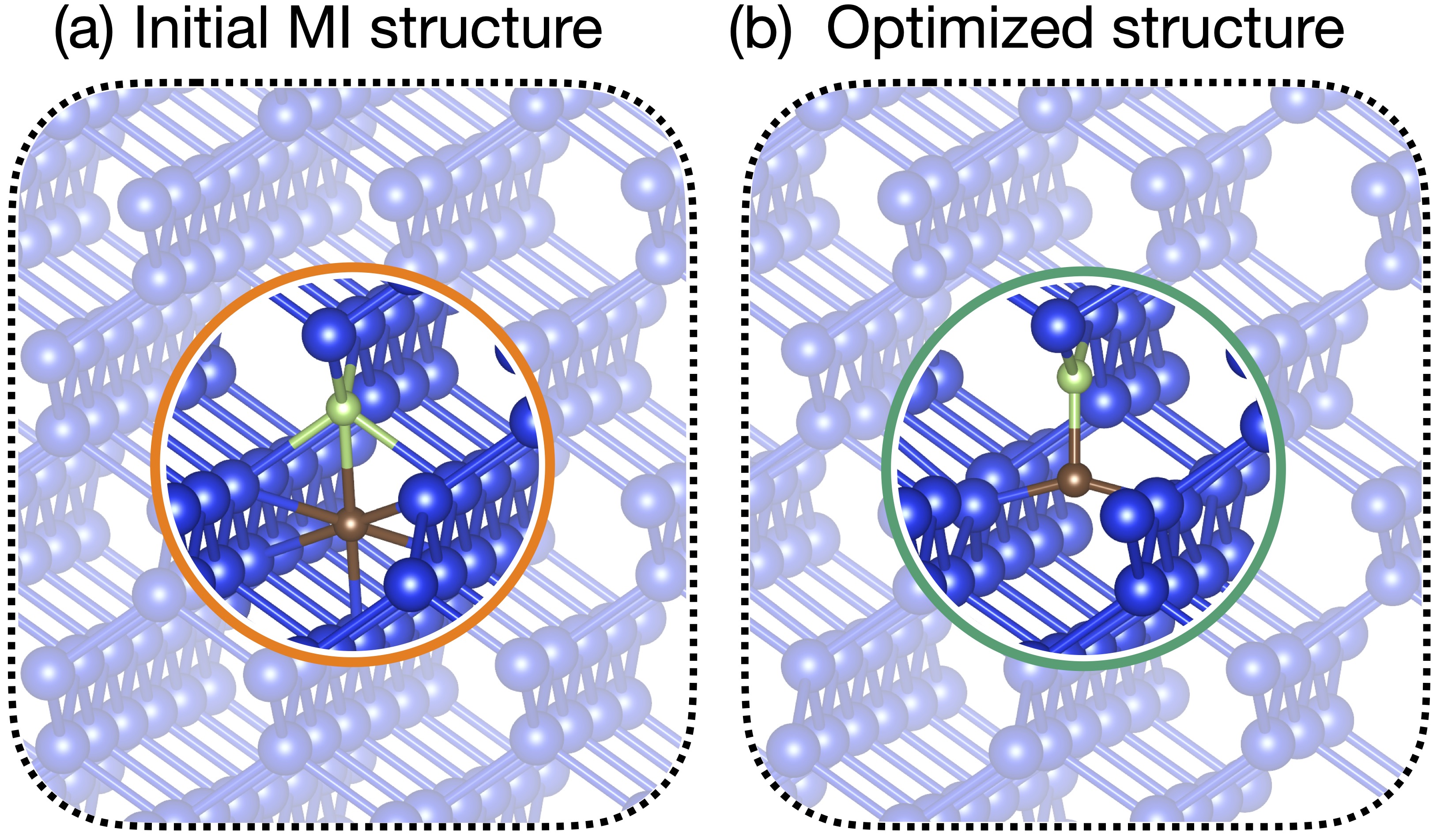}    
 	\caption{(a) One of the initial defect configurations in the MI high-throughput database, with an acceptor substitution on Si, and a carbon with hexagonal interstitial. These two defects are placed as the nearest neighbours. (b) Optimized structure results in T-like defect configuration with $\rm C_{2V}$ symmetry.
 	}
 	\label{Fig.1 init MI defect}
\end{figure}

We also note that the T center-like defects formed through relaxation of the substitutional-interstitial complexes during our high-throughput screening. Interestingly, only the hexagonal interstitial results in the formation of T center-like defect (see Figure \ref{Fig.1 init MI defect}) while all the other combinations do not. The only exception is the (B-C)$_{\rm Si}$, which forms in either case. While hexagonal interstitial in silicon is often a metastable defect \cite{Xiong2023}, our finding highlights the importance of considering metastable defects when building a comprehensive complex quantum defect database as recently also discussed by Deak et al.\cite{Deak2023}.

\subsection{Optical properties}

\begin{table*}[ht!]
    \caption{\label{Tab: defects}Kohn-Sham level difference ($\Delta $KS), zero-phonon line (ZPL), transition dipole moment (TDM), the relevant charge transition level (CTL) with respect to $E_{\rm VBM}$, bound exciton stability (BES), and the Debye-Waller factor (DWF) of the T and T-like centers.}
    \begin{ruledtabular}
    \begin{tabular}{c c c c c c c}
Defect & $\Delta$KS (eV) & ZPL (meV) & TDM (Debye) & CTL (0/$-1$)(eV) & BES (meV) & DWF\\ \hline
 T center & 1.04 & 985 & 0.46 & 1.070 & 85 & 0.165\\
 (B-C)$\rm _{Si}$& 1.121 & 999 & 0.65 & 1.088 & 89 & 0.275\\
 (Al-C)$\rm _{Si}$& 1.029 & 790 & 1.81 & 0.891 & 101 & 0.136\\
 (Ga-C)$\rm _{Si}$& 1.096 & 908 & 1.11 & 1.004 & 96 & 0.198\\
 (In-C)$\rm _{Si}$& 1.104 & 914 & 1.14 & 1.031 & 117 & 0.139\\
 (Tl-C)$\rm _{Si}$& 1.126 & 959 & 0.81 & 1.081 & 122 & 0.116
\end{tabular}
\end{ruledtabular}
\end{table*}

A high performance spin-photon interface requires excellent optical properties, including bright emission at appropriate wavelengths and minimal losses in the phonon sidebands of photoluminescence (i.e., high Debye-Waller factor). Although it is promising that T center-like defects share similar electronic structures with the T center, we performed more precise optical computations to evaluate their performance. We start off by computing the zero-phonon line (ZPL) of the quantum defects. The excited states of the T-like series are prepared using the constrained HSE ($\Delta$-SCF method) by enforcing the occupation of $b_1$ state in the spin-down channel an creating a hole at the VBM \cite{Gali2009,Dhaliah2022}. The ZPL is the energy difference between the relaxed defect in the excited and ground state. We also estimate the brightness or radiative lifetime by the transition dipole moment obtained from the ground-state wavefunctions, as detailed in the methods section \cite{Gali2009,Alkauskas2016.PRB,Davidsson2020}. These results are summarized in Table~\ref{Tab: defects}. We found that all of the T-like series emit in the near IR region, with Al-C and B-C complex exhibiting the lowest and highest ZPL energies at 790 meV and 999 meV, respectively. Given that all these defects are excited through transition between the host and defect bands, forming a defect-bound exciton, it is crucial to consider their stability of this exciton. For the T center-like defects, the bound exciton stability measures the tendency for the bound exciton to ``break'' and form a free hole and the negatively charged defect, offering a measure of the defect's stability against photoionization rather than optical excitation. This is effectively the energy difference the ZPL and the (0/$-1$) charge transition level (see next section), as shown in Table~\ref{Tab: defects}. Generally, the T-like series show bound exciton stability that is similar to the T center, ranging between 89 meV to 122 meV. This indicates a stable bound exciton (at low temperature) given the positive binding energies. We note that the computed bound exciton binding energy is overestimated by first-principles computations for the T center probably due to the slow convergence with respect to the supercell size.\cite{Dhaliah2022}. In any case, our analysis supports the stability of T center-like bound excitons at sufficiently low temperatures for readout and initialization. Typically, the T center operates at a few Kelvin \cite{Higginbottom2022}, well below the temperature that will break the defect-bound exciton.

Under the the Huang-Rhys approximation\cite{Alkauskas2014}, we now compute the Debye-Waller factor (DWF) that quantifies the proportion of photons emitted into the ZPL as opposed to the phonon sideband which is not suitable for quantum operations. We first benchmark our results by comparing the computed DWF of the T center to the experimental measurements. Our computed DWF of 0.165 aligns reasonably with the experimental value of 0.23 reported in Ref.\cite{Bergeron2020}. See the details of the computed PL in Supplementary Note 2. Previous studies have reported an overestimated DWF of 0.856 for the T center, which might attribute to the finite size effects in small supercells or specific technical details in the $\Delta$SCF computations \cite{Ivanov2022}. Combining DWF (efficiency) with the TDM (brightness) allows us to comprehensively assess the optical properties of the T-like defects. Notably, We found that B-C emerges as closely analogous to the T center, with a predicted ZPL of 999 meV -- near the T center's 985 meV -- and comparable brightness values (TDM of 0.65 D for B-C and 0.46 D for the T center). However, B-C distinguishes itself with a DWF of 0.275, approximately 1.7 times that of the T center, marking the highest efficiency within the T-like series. We note that it has been recently suggested that a higher DWF not only enhances the emission efficiency of the ZPL but also reduces the non-radiative recombination, which plays a predominant role in the telecom wavelength, as detailed in Ref.\cite{Turiansky.arxiv.2024}. According to their model, we estimated that B-C's non-radiative recombination rate is around three orders of magnitude lower than that of the T center, positioning B-C as a highly promising spin-photon interface with superior emission efficiency at a wavelength comparable to the T center. 

The other defects among the series also present interesting properties that are worth noting. Al-C, in particular shows the highest transition dipole moment in the series with 1.81 D. While emitting at a longer wavelength of 790 meV, it is estimated to emit with a radiative lifetime of 0.7 $\mu$s, which is an order of magnitude smaller than the T center (7.8 $\mu$s). A similar trend is observed with Ga-C, In-C, and Tl-C, where their radiative lifetimes are 3 to 5 times shorter than that of the T center, with values of at 1.7, 1.6, and 2.72 $\mu$s, respectively. Meanwhile, their DWFs are found to be comparable to that of the T center. We observe a general trend where the presence of heavier elements in the defects tends to reduce the DWF (with the exception of Al-C), possibly due to the greater amount of relaxation during excitation the ($\Delta$Q). To further validate the above findings, we compared the transition dipole moment and vertical excitation energies (excited electronic structures with ground-state atomic structures) calculated using both $\Delta$SCF and time-dependent DFT (TDDFT) methods, both on top of HSE. The results were consistent across both methodologies, with detailed comparisons available in Supplementary Note 3. We note that the recent development of the TDDFT method in periodic systems has enabled the evaluation of nuclear forces\cite{Jin.JCTC.2023}. However, this topic is beyond the scope of our current work.

\subsection{Comparison with experiments}

\begin{figure}[t]
 	\centering
 	\includegraphics[width=0.45\textwidth]{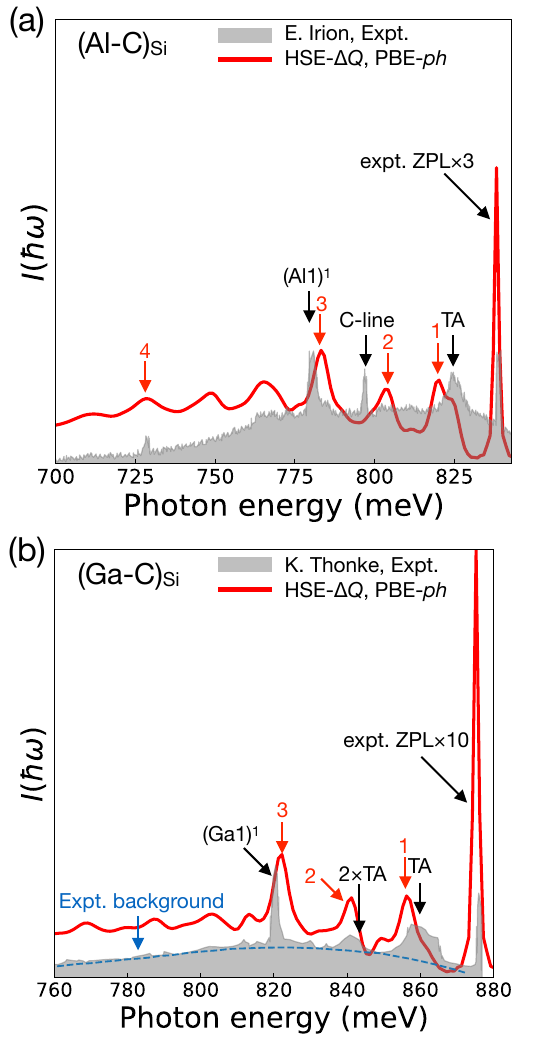}    
 	\caption{Comparison between the simulated PL spectrum and the experimental measurements for (a) AlC and (b) GaC complex defects. The characteristic peaks are highlighted and indexed for the simulated PL lineshape.
 	}
 	\label{Fig.6 PL of AlC and GaC}
\end{figure}

There is a track-record of studies on color centers in silicon preceding the advent of QIS. There is substantial evidence from this past literature suggesting that the complexes we identified exist and have been synthesized previously. Photoluminescence (PL) spectra of Al-C and Ga-C complexes have been reported. Irion and colleagues observed the ``Al1'' spectrum in electron-irradiated Al-doped silicon, with a measured ZPL of 836 meV, closely aligning with our predicted value of 790 meV \cite{Irion.APA.1989}. The comparison between the computed and experimental PL lineshapes, as shown in Figure~\ref{Fig.6 PL of AlC and GaC}(a), reveals a good correspondence. It is particularly notable that the signature ``(Al1)$^1$'' peak around 780 meV is captured by our model's peak 3. This quasi-localized vibrational mode (LVM) is shown in Suppelemtary Note 1 and involves a fair amount silicon around the defect. This is in correspondence to the experimental interpretation that this mode is a defect-meditated zone-center optical phonon mode (O$^{\Gamma}$)\cite{Irion.APA.1989}. We note that the ``C-line'' at 789 meV shown in the experimental data is associated with carbon-oxygen complex\cite{Thonke.JPC.1985}. Overall, the agreement between theory and experiment indicates that the experimentally reported defect is likely to be the T center-like Al-C. Similarly, Thonke and the coauthors reported ``Ga1'' spectrum from Ga-doped silicon placed under electron irradiation with a subsequent annealing \cite{Thonke.PRB.1985}. The reported ZPL of the Ga1-line at 875 meV again closely matches our predicted value of 908 meV. Figure~\ref{Fig.6 PL of AlC and GaC}(b) plots the measured and the predicted PL spectrum that are in excellent agreement. The characteristic (Ga1)$^1$ peak at 820 meV that involves the quasi-localized mode of the defect is reproduced by the peak 3 in the simulated PL. Compared to (Al1)$^1$, (Ga1)$^1$ shows a stronger localization character, as detailed in Supplementary Note 2. Our results strongly suggest the existence of Ga-C pair complex defects as the origin of the Ga1 line.
We note that both the gallium and aluminium experimental data indicates a Zeeman splitting in the emission spectra. This is consistent with the doublet nature of the T center like defects \cite{Irion.APA.1989,Thonke.PRB.1985}.

Regarding the B-C complex, clear PL evidence remains elusive. We note that there was early experimental confusion of whether boron-carbon complex could be responsible for the T-line \cite{Irion.JPC.1985}. Though experimental efforts later resolve the confusion by identifying the defect symmetry of the T center to be $C_{1h}$ though stress measurements \cite{Safonov1996}. Interestingly, our findings suggest that the PL lineshapes of the B-C complex closely resemble those of the T center. The higher DWF of B-C complex might result in its sideband being covered by the phonon sideband of the T center, see Supplementary Note 4. We hypothesized that low concentration of B-C could coexist with the T center in B-doped silicon samples. Without a targeted synthesis, it might be challenging to isolate these two defects. Recently, the G and G$^*$ centers have suggested that defects with similar PL signature can be confused for each other\cite{Redjem2020,Komza2022arXiv,Durand.arxiv.2024}. Finally, to the best of our knowledge, there is no clear experimental evidence showing the existence of (In-C)$_{\rm Si}$ and (Tl-C)$_{\rm Si}$ complexes \cite{Sauer.PBC.1983,Jones.JAP.1981,Weber.JL.1981}. Their synthesizability might be limited by the thermodynamic factors which will be discussed in greater details in the subsequent section.

\subsection{Thermodynamic analysis}
\label{sec:thermo}

\begin{figure*}[t]
 	\centering
 	\includegraphics[width=0.9\textwidth]{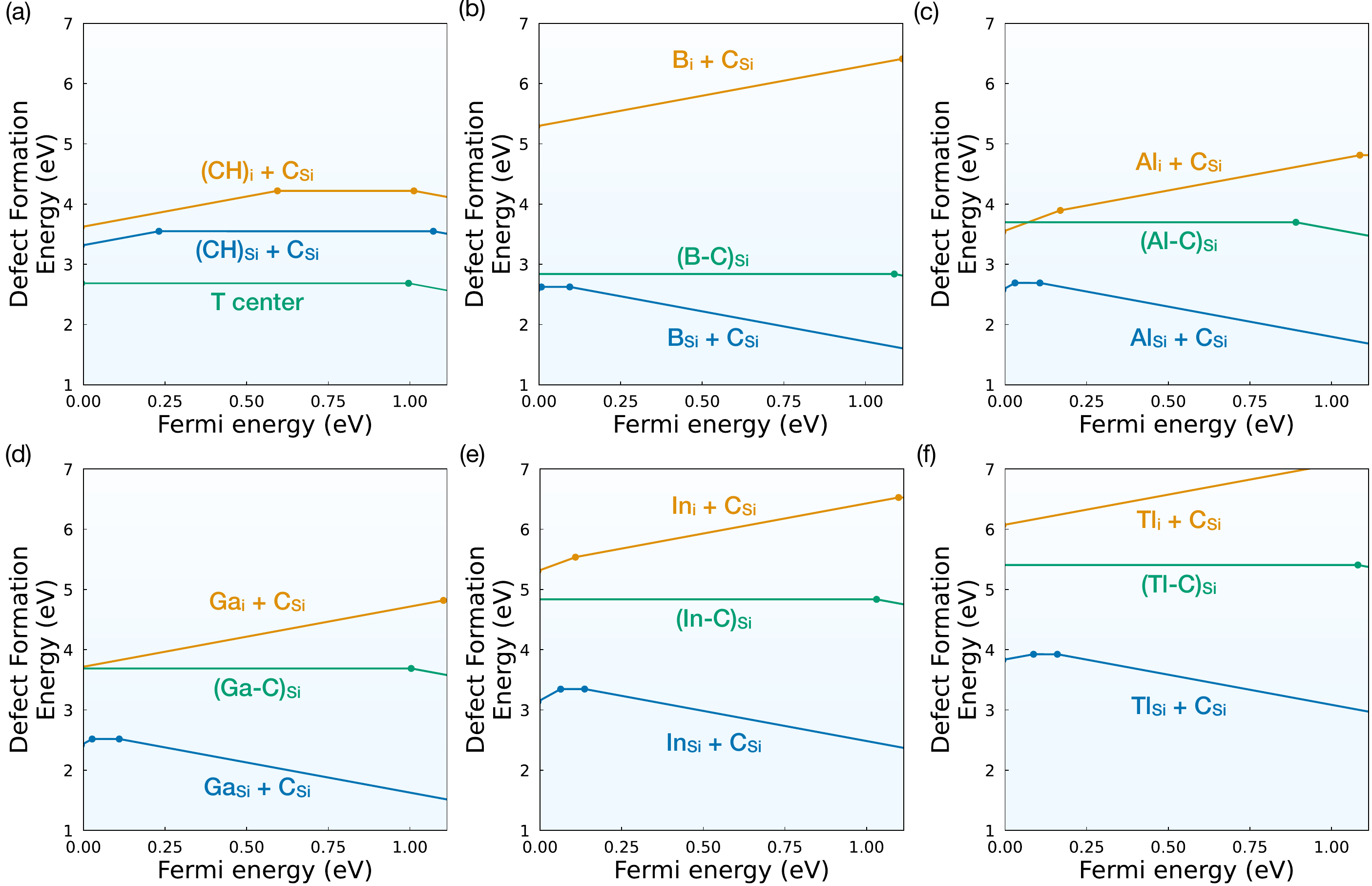}    
 	\caption{Defect formation energy vs. the Fermi energy for (a) T and (b-f) T-like defects. The zero of Fermi energy is referenced to the valence band maximum. Elemental chemical potentials are used as reference. Decomposition into simple defects are considered in this case, including C$\rm _{Si}$ and CH/acceptor interstitial (CH$\rm _i$/A$\rm _i$).
 	}
 	\label{Fig.4 formation}
\end{figure*}

We have identified the T center-like defects as promising candidates for spin-photon interfaces. While previous experimental PL have shown evidences for the existence of these defects, we perform now a thermodynamic analysis to further investigate their stability and suggests pathways for their formation. Figure~\ref{Fig.4 formation} shows the defect formation energies with respect to the Fermi level for all T center-like defects. All of the defects show similar charge stability with the neutral charge dominate the band gap. In fact, most of the T-like defects, similar to the T center, show (0/$-1$) charge transition level around 1 eV, with Al-C shows a slightly lower level at 0.89 eV, as summarized in Table~\ref{Tab: defects}. Using first-principles computations, we investigate the potential decomposition of the complex defect centers into simpler defects. The chemical potentials used for evaluating formation energies are referenced to the elemental forms. Previous studies suggested that the T center forms by a C$_{\rm Si}$ that captures an interstitial CH$_{\rm i}$\cite{Safonov1996}. Based on this insight and our previous work on the T center, we evaluate both C$_{\rm Si}$ + CH$_{\rm i}$/A$_{\rm i}$ and C$_{\rm Si}$ + CH$_{\rm C}$/A$_{\rm C}$ as competing products with the complex. Our results confirms that both the T and the T centers-like defects are more stable than the C$_{\rm Si}$ + CH$_{\rm i}$/A$_{\rm i}$. This supports the hypothesis that they could form by a substitutional defect trapping an interstitial. Furthermore, our analysis indicates that all T center-like defects tend to thermodynamically decompose into substitutional defects. However, it is likely that the complex defects could be kinetically trapped. This is because the decomposition into substitutional defects is a complex rearrangement involving a silicon vacancy, which could be kinetically hindered at lower temperatures. The experimental evidence for the existence of (Ga-C)$_{\rm Si}$ and (Al-C)$_{\rm Si}$ outlined in the previous section supports this view. We also note that other complex in silicon such as the G-center have also been shown to be kinetically stabilized\cite{Deak.Nat.Commun.2023}. Additionally, we also performed \textit{ab initio} molecular dynamics simulations at 300 K for each of the defects. Over a 50 ps production run, no dissociation events were observed (see Supplementary Note 5 for more details).

\begin{figure*}[t]
 	\centering
 	\includegraphics[width=0.9\textwidth]{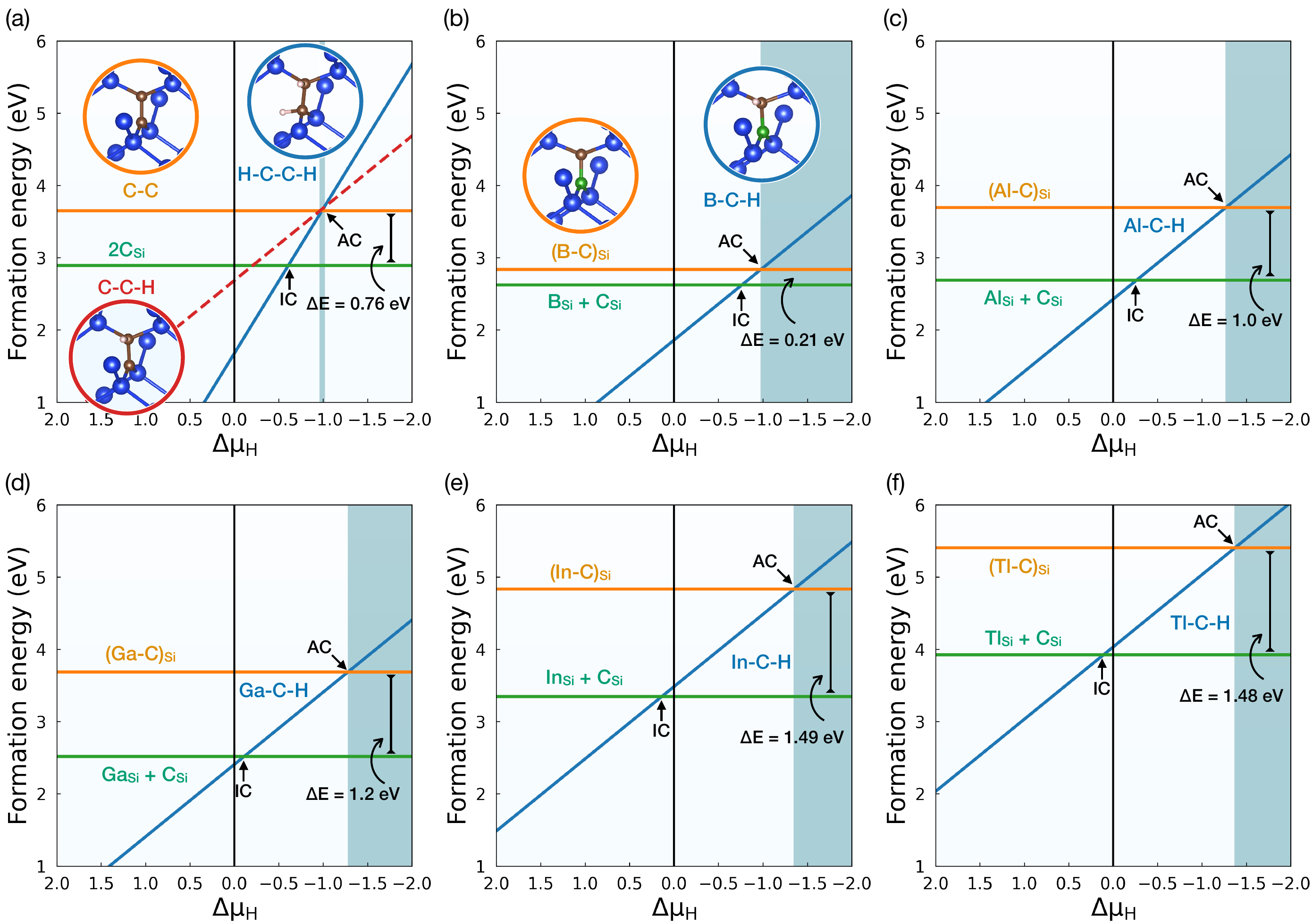}
        \caption{Stability of the T and T-like defects in a grand canonical ensemble open to hydrogen. The hydrogenated defect structures are shown in the inset circles. Chemical potential of hydrogen corresponds to implantation (IC) and annealing condition (AC) are shown by the arrow. The annealing hydrogen poor chemical potential are highlighted by the shaded area. 
 	}
 	\label{Fig.5 GC}
\end{figure*}

Both our previous first-principles work and experimental evidences have shown that controlling the hydrogen content or chemical potential of hydrogen ($\mu_{\rm H}$) is essential to the production of the T center\cite{Dhaliah2022,Lightowlers1994,MacQuarrie2021}. We now extend this investigation to the formation of the T center-like defects. Since the T center-like series also involve unpaired electrons on the carbon atoms (i.e., a radical in chemistry), we expect that passivating this unpaired electron using hydrogen might favor the formation of the T center-like defects. Given hydrogen's high mobility, we suggest a two-step process for creating these defects (with a high yield): 1) synthesizing hydrogenated T center-like defects (i.e., (A-C-H)$_{\rm Si}$), and 2) selectively removing hydrogen while the other elements are kinetically frozen to lead to the formation of paramagnetic defects (i.e., (A-C)$_{\rm Si}$). We now user first principles computed energies and consider a system open to hydrogen with chemical potential $\mu_{\rm H}$, where $\mu_{\rm H}$ = 0 eV is referenced to the H$_2$ molecule at 0 K. In practice $\mu_{\rm H}$ is controlled by hydrogen gas or water partial pressure and temperature. The grand-canonical thermodynamic potential of different defects in function of the hydrogen chemical potential is shown in Figure ~\ref{Fig.5 GC} for the T center and T center-like defects. Figure ~\ref{Fig.5 GC}(a) focuses on the T center where the the lines are respectively indicative of the thermodynamic potential for the fully hydrogenated T center (H-C-C-H) in blue, fully dehydrogenated (C-C) in orange and partially hydrogenated (C-C-H) in dashed red. The single carbon substitution (C$\rm_{Si}$) competing with the complex is shown in green. Our previous work has shown that the T center is first synthesized at high $\mu_{\rm H}$ (ion implantation followed by immersing in boiling water) that favors the formation of H-C-C-H over the decomposition into substitutional defects (green line) \cite{Dhaliah2022}. The complex is then dehydrogenated by carefully controlling the $\mu_{\rm H}$ during an annealing step. Notably, the T center is very sensitive to this annealing step and is only stable in a very narrow window (blue area in ~\ref{Fig.5 GC} (a)). A too high $\mu_{\rm H}$ will not dehydrogenate the H-C-C-H complexes at all and a excessively low $\mu_{\rm H}$ will form fully dehydrogenated defects. This analysis underscores the sensitivity required for successfully processing the T center.

We now carry out the same analysis for the T center-like defects, as shown in Figure~\ref{Fig.5 GC} (b) to (f). Similar to the T center, a high $\mu_{\rm H}$ promotes the hydrogenated defect (blue line), even when competing against the decomposition into substitutional defects (green line). This suggests that ion implantation with hydrogen might be favoring the formation of T center-like (hydrogenated) defects. To then produce the paramagnetic, dehydrogenated form of T center-like defects, we propose annealing the system at a low $\mu_{\rm H}$. Assuming this is done under the conditions that only allow hydrogen to diffuse, we would be able to kinetically stabilize the acceptor-carbon complex. This mechanism is applicable for all T-center like defects. What changes is the conditions necessary to achieve the formation of the hydrogenated defect and to anneal out the hydrogen. The upper (lower) bound of $\mu_{\rm H}$ for ion-implantation IC (annealing condition AC) are presented in Table~\ref{Tab2: ICAC}. For instance, ion-implantation of the T center needs to be performed with $\mu_{\rm H}$ more positive than $-0.61$ eV, corresponds to a temperature lower than 624$^{\circ}$C in hydrogen atmosphere ($p{\rm H_2}$=1 bar), which can be easily fulfilled. Due to the narrow annealing range with $\mu_{\rm H} \sim -0.99$ eV, the T center needs to be annealed around 462$^{\circ}$ by assuming the $p{\rm H_2}$=1e$-7$ bar, in reasonable agreement with experimental temperature of 400$^{\circ}$ in nitrogen atmosphere. Comparatively speaking, B-C indicate a superior synthesizability with a wider range of ion-implantation and annealing temperature. Ion-implantation can be performed as long as $T^{\rm IC} < 856 ^\circ$C, and with annealing $T^{\rm AC} > 440 ^\circ$C under the same conditions of the T center. In general, heavier group III complex becomes more difficult to synthesize as their $\mu_{\rm H}^{\rm IC}$ becomes more positive (lower $T^{\rm IC}$) and their $\mu_{\rm H}^{\rm AC}$ becomes more negative (higher $T^{\rm AC}$). Additionally, the decomposition energy $\Delta E_{d}$ that measures the decomposition tendency of the acceptor-carbon complex into simple defect counterparts show the similar trend, as summarized in Table~\ref{Tab2: ICAC}. Therefore, the heavier group III complex could become more difficult to be kinetically stabilized due the decomposition tendency, and this might explain the lack of experimental evidence of In-C and Tl-C in silicon.

\begin{table}[ht!]
    \caption{\label{Tab2: ICAC} Chemical potential of hydrogen for ion-implantation and annealing condition $\mu_{\rm H}^{\rm IC}$/$\mu_{\rm H}^{\rm AC}$ (eV) for the T and T-like centers, their corresponding temperature T$^{\rm IC}$/ T$^{\rm AC}$ ($^{\circ}$C), and the decomposition energy $\Delta E_{d}$ (eV).}
    \begin{ruledtabular}
    \begin{tabular}{c c c c c c c}
Defect & $\mu_{\rm H}^{\rm IC}$ & $\mu_{\rm H}^{\rm AC}$ & $\Delta E_{d}$\\ \hline
 T center         & $-$0.61 & $-$0.99  & 0.76 \\
 (B-C)$\rm_{Si}$  & $-$0.77 & $-$0.98  & 0.21 \\
 (Al-C)$\rm _{Si}$& $-$0.26 & $-$1.27  & 1.00 \\
 (Ga-C)$\rm _{Si}$& $-$0.11 & $-$1.28  & 1.20 \\
 (In-C)$\rm _{Si}$& 0.14  & $-$0.52  & 1.49 \\
 (Tl-C)$\rm _{Si}$& 0.03  & 0.11   & 1.48 
\end{tabular}
\end{ruledtabular}
\end{table}

\section{Discussion}

Using a high-throughput computational screening approach, we identified a series of new quantum defects analogous to the T center in silicon. These defects comprise of an acceptor-carbon complex on the silicon site, exhibiting similar electronic structure to the T center where the unpaired electron on the carbon $p$ orbital sets the in-gap defect level close to the CBM, leading to telecom emission. We propose that these T center-like defects could be used as spin-photon interfaces similarly to the T center, and be excited through a defect-bound exciton. Experimental evidences show that some of these defects have probably been synthesized and reported as color center in silicon in previous studies (well before the advent of quantum technologies). 

This series of quantum emitters offers a range of wavelength such as telecom $O$-band (B-C), $S$-band (Al-C), and $E$-band (Ga-C). Many of these defects have a radiative lifetime similar to the T center with the exception of Al-C complex whose lifetime is an order of magnitude shorter. The T center is typically operated at a few Kelvin and will not be able to be used at higher temperature because of the low bound exciton energy. We expect similar operational temperature range for these T center-like color centers. The T center has shown spin coherence time in the millisecond range (T$_2$=2.1 ms)\cite{Bergeron2020} and we expect the T center-like to be comparable, as spin coherence is predominately influenced by the host nuclei spin bath at these low temperatures. 

All the T center-like defects have an important practical advantage over the T center due to their higher symmetry (C$_{2v}$ vs C$_{1h}$). This benefits the control of these defects in terms of alignment with the magnetic field needed to split the doublet state. Considering the crystal axis, an ensemble of the T center can consist up to 12 inversion symmetry orientations, each responsible for a branch of the optically detectable magnetic resonance (ODMR) measurements. In contrast, it is expected that each of the T-like defects will show only 6 branches in an ensemble, simplifying the identification of defect orientation during experiments. This information facilitates the alignment of the magnetic field for optimal spin readout\cite{Sukachev.PRL.2017,Raha.NC.2020} and integration with nanophotonic structures. In terms of synthesis, we propose that a process including the synthesis of hydrogenated T center-like followed by an annealing step could lead to a high yield of the T center-like defects. Our analysis indicates that the annealing might not require as precise control as for the T center, presenting a potential advantage for these defects in terms of defect production yield. While all the T center-like defects are of interest, we note that B-C is especially appealing as it emits in the O-band similarly to the T center, with a comparable radiative lifetime and a significantly higher DWF.

\section{Conclusion}

Silicon color centers are strong contenders as qubits in scalable quantum networking and computing. Exploring the complex defect database of more than 22,000 defects that are built by our automated high-throughput computational workflow has led to the discovery of a class of spin-photon interface based on a group III element bound to carbon on a silicon site. These defects are analogous structurally and electronically to the T center and show attractive properties including a doublet ground state and efficient emission of photons in the telecom. Compared to the T center, these new defects have higher symmetry facilitating the alignment with the magnetic field during ODMR and can have more attractive optical properties. While experimental evidences show that these color centers can be produced, we suggest a synthesis route that could enhance the yield of the T-like defects by forming hydrogenated complexes that are subsequently dehydrogenated through annealing. Our work highlights the importance of comprehensive complex defects database for discovery of new spin-photon interface and motivates further experimental and theoretical investigations of these T center-like defects for quantum applications.

\section{Methods}

The high-throughput defect computations were performed using the automatic workflows that are implemented in atomate software package~\cite{Jain2013, Mathew2017,Ong2013}. The first-principles calculations were performed using Vienna Ab-initio Simulation Package (VASP)~\cite{G.Kresse-PRB96,G.Kresse-CMS96} with the projector augmented wave (PAW) method~\cite{P.E.Blochl-PRB94}. All the calculations were spin-polarized with the Perdew-Burke-Erzhenhoff (PBE) functional. Charged defects were simulated using a supercell size of 216 atoms. 520 eV cutoff energies were used for the plane-wave basis and only the $\Gamma$ point was used to sample the Brillouin zone. All the defect structures were optimized at a fixed volume until the ionic forces are smaller than 0.01~eV/$\AA$. On top of these PBE computations, we have performed single-shot Heyd-Scuseria-Ernzerhof (HSE) (HSE$_0$) computations that have shown to correct the PBE electronic structure at a minimal cost \cite{Xiong2023, Heyd2003}. In total our database includes 22266 complex charged defects from 56 elements that are identified as easily ion-implantable\cite{Xiong2023}. Each of these complex defect was initialized by a substitution and a interstitial (tetrahedral or hexagonal) defects.

For the T-like defects, we performed lower throughput more precise computations. The full self-concistent HSE was used in a larger 512 atoms supercell with a reduced cutoff of 400 eV. For all the defect calculations, the input generation and output analysis are performed using PyCDT~\cite{Broberg2018}. The formation energy of each charged defect reads:

\begin{equation}
    E_\mathrm{form}[X^q] = E_\mathrm{tot}[X^q] - E_\mathrm{tot}^\mathrm{bulk} - \sum n_i \mu_i + q E_f + E_\mathrm{corr}
\label{eq:defect}
\end{equation}

where the formation energy is expressed as a function of the Fermi level $E_f$~\cite{Zhang1991,Komsa2012}. $E_\mathrm{tot}[X^q]$ and $E_\mathrm{tot}^\mathrm{bulk}$ are the total energies of the defect-containing supercell (for a given defect $X$ in the charge state $q$) and the bulk, respectively. The third term represents the energy needed to exchange atoms with thermodynamic reservoirs where $n_i$ indicates the number of atoms of species $i$ removed or added to create the defect, and $\mu_i$ is their corresponding chemical potential. In this work, we used elemental chemical potentials as the reference. The fourth term represents the energy to exchange electrons with the host material through the electronic chemical potential given by the Fermi level. Finally, the last term is a correction accounting for spurious image-charge Coulomb interactions due to finite supercell size, as well as potential-alignment corrections to restore the position of the bulk valence band maximum (VBM) in charged-defect calculations due to the presence of the compensating background charge density~\cite{Freysoldt2011,Kumagai2014}. Kumagai correction was applied throughout this work~\cite{Kumagai2014}.

For the grand canonical thermodynamic analysis that opens to hydrogen, the chemical potential of hydrogen $\mu_{\rm H}$ can be linked to the synthesis conditions using an ideal gas model.

\begin{equation}
{\mu}_{\rm H}(T,P) = \frac{1}{2}(E_{\rm H_2} - TS_{\rm H_2}^{\rm exp} + RT{\rm ln}(P_{\rm H2})), 
\end{equation}
where $E_{\rm H_2}$ is the energy of the H$_2$ molecule. $S_{\rm H_2}^{\rm exp}$ is the entropy of H$_2$ measured experimentally at standard condition~\cite{Chase1998}, $p_{\rm H2}$ is the partial pressure of H$_2$. We used the H$_2$ molecule at 0K as our reference, zero chemical potential.

The transition dipole moment was evaluated using the single-particle wavefunction that was calculated at HSE level and further processed with PyVaspwfc code \cite{Zheng2018}. The radiative lifetime was approximated using Wigner-Weisskopf theory of fluorescence: \cite{Gali2019,Alkauskas2016.PRB,Davidsson2020}

\begin{equation}
\frac{1}{\tau}=\frac{n_{r} (2 \pi)^3 \nu^{3}|{ \boldsymbol{\mu}}|^{2}}{3 \varepsilon_{0} h c^{3}},
\end{equation}
where $\tau$ is the radiative lifetime, $n_{r}$ is the refractive index of the host, $\nu$ is the transition frequency corresponding to the energy difference of the Kohn-Sham levels, ${\boldsymbol{\mu}}$ is the transition dipole moment, $\varepsilon_{0}$ is the vacuum permittivity, $h$ is the Planck constant, and $c$ is the speed of light. The transition dipole moment is given as:

\begin{equation}
\boldsymbol{\mu}_{k}=\frac{\mathrm{i} \hbar}{\left(\epsilon_{\mathrm{f}, k}-\epsilon_{\mathrm{i}, k}\right) m}\left\langle\psi_{\mathrm{f}, k}|\mathbf{p}| \psi_{\mathrm{i}, k}\right\rangle,
\end{equation}
where $\epsilon_{\mathrm{i}, k}$ and $\epsilon_{\mathrm{f}, k}$ are the eigenvalues of the initial and final states, $m$ is the electron mass, $\psi_{\mathrm{i}}$ and $\psi_{\mathrm{f}}$ are the initial and final wavefunctions, and $\mathbf{p}$ is the momentum operator.

Time-dependent density functional perturbation theory was used to compute transition dipole moment and vertical excitation energies in the CP2K package using the Gaussian Plane Waves method\cite{cp2k,cp2k_tddft}. The Gaussian type orbitals are mapped to the multigrid solver with a relative cutoff of 50 Ry. A TZVP basis was used when available, otherwise the DZVP basis was used, along with an auxiliary basis to accelerate hybrid computations\cite{cp2k_basis1,cp2k_basis2}. The hybrid functional HSE06 was used in the exchange-correlation kernel for the TDDFT calculations. In this work used GTH pseudopotentials.\cite{cp2k_pseudopotential}. Only the $\Gamma$ point was used to sample the Brillouin zone. All structures were relaxed in fixed volume in CP2K until the magnitude of forces was less than 0.01 eV/\r{A}. The convergence of the excited state energies was set to 1e$-$7 eV. 

PL spectra were computed with ground and excited states structures at HSE level and phonon calculated within GGA level, following the method proposed by Alkauskas \cite{Alkauskas2014}. All the calculations were performed in a 512-atoms cell. Ground-state phonon is assumed to be identical to the excited states~\cite{Razinkovas2021}. The phonon properties are computed using \texttt{Phonopy} and compressive-sensing method implemented in \cite{phononpy,pheasy,Zheng2024arxiv}. The PL spectra were plotted using \texttt{pyphontonics} with a Gaussian broadening of 2 meV\cite{pyphotonics}. The ZPL positions were set to the values that were computed using HSE $\Delta$SCF method.

\section{Acknowledgments}
\begin{acknowledgments}
This work was supported by the U.S. Department of Energy, Office of Science, Basic Energy Sciences 
in Quantum Information Science under Award Number DE-SC0022289. 
This research used resources of the National Energy Research Scientific Computing Center, 
a DOE Office of Science User Facility supported by the Office of Science of the U.S.\ Department of Energy 
under Contract No.\ DE-AC02-05CH11231 using NERSC award BES-ERCAP0020966. 
\end{acknowledgments}

\end{document}